\documentclass[10pt,twocolumn,aps,prc,longbibliography,nofootinbib,floatfix]{revtex4-1}
\usepackage{graphicx,bm,xcolor, bbold}
\usepackage[normalem]{ulem}
\usepackage{hyperref}
\hypersetup{colorlinks, linkcolor={blue},citecolor={blue},urlcolor={blue}}
\usepackage{graphicx}
\usepackage{amsmath,amssymb,amsfonts}
\usepackage{bm}
\usepackage{color}
\graphicspath{fig}
\usepackage{epstopdf}
\usepackage{multirow}
\usepackage{scalerel}

\newcommand{\paral}{\stretchrel*{\parallel}{\perp}}

\newcommand{\vect}[1]{\bm{\mathrm{#1}}}

\newcommand{\Jv}{\vect{J}}
\newcommand{\av}{\vect{a}}
\newcommand{\bv}{\vect{b}}
\newcommand{\jv}{\vect{j}}
\newcommand{\kv}{\vect{k}}
\newcommand{\nablav}{\vect{\nabla}}
\newcommand{\pv}{\vect{p}}

\newcommand{\Qv}{\vect{Q}}
\newcommand{\rv}{\vect{r}}
\newcommand{\rhov}{\vect{\rho}}
\newcommand{\vv}{\vect{v}}
\newcommand{\vvbar}{\bar{\vect{v}}{}}
\newcommand{\xv}{\vect{x}}
\newcommand{\xvhat}{\hat{\vect{x}}}
\newcommand{\yvhat}{\hat{\vect{y}}}
\newcommand{\zvhat}{\hat{\vect{z}}}
\newcommand{\lambdavpar}{\vect{\lambda}_{\paral}}
\newcommand{\sumpbppar}{\sum_{\pv_b,p_{\paral}}}
\newcommand{\calH}{\mathcal{H}}

\newcommand{\nv}{\vect{n}}

\begin{document}
\title{Superfluid fraction in the rod phase of the inner crust of neutron stars}
\author{Giorgio Almirante}
\email{giorgio.almirante@ijclab.in2p3.fr}
\affiliation{Universit\'e Paris-Saclay, CNRS/IN2P3, IJCLab, 91405 Orsay, France}
\author{Michael Urban}
\email{michael.urban@ijclab.in2p3.fr}
\affiliation{Universit\'e Paris-Saclay, CNRS/IN2P3, IJCLab, 91405 Orsay, France}
\begin{abstract}
The rod phase as it is expected in the bottom layers of neutron-star crusts is analyzed within the Hartree-Fock-Bogoliubov framework. In order to well describe the interplay between band structure and superfluidity, periodicity of the lattice is taken into account using Bloch boundary conditions. A relative flow between the rods and the surrounding neutron gas is introduced in a time-independent way. This induces a non-trivial phase of the complex order parameter, leading to a counterflow between neutrons inside and outside the rods. With the resulting current, we compute the actual neutron superfluid fraction. For the latter our results are significantly larger than previous ones obtained in normal band theory, indicating that the normal band theory overestimates the entrainment effect.
\end{abstract}
\maketitle
%
\section{Introduction}
In the inner crust of neutron stars, protons and neutrons cluster in very neutron-rich structures. These clusters are probably arranged in a periodic lattice due to the interplay between the short-range nuclear force and the long-range Coulomb interaction \cite{Negele73,Martin15,Thi21}. In addition, there is a gas of unbound neutrons at densities where pure neutron matter is superfluid, together with a background degenerate relativistic electron gas which ensures charge neutrality and $\beta$-equilibrium \cite{Chamel08}. The superfluid component of the crust could have observable consequences for the hydrodynamical and thermodynamical properties of the star \cite{Page12}. It is also the main source of uncertainty in determining quantities such as shear modes \cite{Tews17}. Moreover, there are mechanisms to explain pulsar glitches which rely on the superfluid component of the crust \cite{Prix02,Carter06A}. But, in order to compare them with observations, one needs to know some microscopic features of the inner crust \cite{Antonelli22}, such as the neutron superfluid fraction.

In this paper, we will only consider the zero-temperature case. Then, in a uniform superfluid, the superfluid fraction is usually equal to $100\%$~\cite{Leggett98}. Interestingly, this is so even in weakly coupled fermionic systems, although only the particles near the Fermi surface are affected by pairing. This can be understood from the fact that the superfluid density is actually not a real density of particles, but a long-wavelength limit of a response function, and therefore involves only particles near the Fermi surface (see \cite{Schrieffer}, p. 218). However, once the system becomes non-uniform, as it is the case in the neutron-star crust, the superfluid fraction gets reduced \cite{Leggett98}.

To compute the actual superfluid density $\rho_S$, the crucial point is the evaluation of the so-called ``entrainment'', that is a non-dissipative force between the superfluid component and the nuclear lattice \cite{Prix02}. In fact, the entrainment concerns all kinds of two-component systems with at least one superfluid part. In the presence of a periodic lattice, a way to understand this kind of effect is the Bragg scattering, in our case of dripped neutrons by the nuclear lattice \cite{Chamel12A}, which is the analog of coherent electron scattering in ordinary solids giving rise to the band gaps \cite{Ashcroft}. In order to do this, one has to deal with complicated band-structure calculations for the neutrons \cite{Chamel05,Chamel06,Chamel12A,Kashiwaba19,Sekizawa22}, and results indicate that the entrainment can be very strong, reducing drastically the superfluid fraction. This is in contradiction with the observed glitches of certain pulsars \cite{Chamel13}, unless one gives up the common belief that only the crust is responsible for the glitches \cite{Andersson12}.

But band theory is not the only way. In Ref.~\cite{Martin16}, the superfluid fraction was computed assuming an irrotational flow in a schematic density profile with simple boundary conditions between the clusters and the neutron gas (we use the word ``cluster'' also for the rods and slabs in the pasta phases). However, results were not in agreement with those obtained previously within the band-theory framework. The entrainment obtained in the hydrodynamical approach is much weaker, and the corresponding superfluid fraction much larger, which if it was true would help to understand the observed Vela glitches. 

Both these approaches have some shortcomings. On the one hand, for the hydrodynamical approach \cite{Martin16} to be valid, one needs Cooper pairs smaller than the spacing of the periodic lattice and the size of the clusters, and unfortunately this is not true in the inner crust of neutron stars. On the other hand, band theory calculations \cite{Chamel05,Chamel06,Chamel12A}, even if pairing is included in the BCS approximation \cite{Carter05A,Chamel10}, are missing the dynamics of the superfluid order parameter. This means that they cannot reproduce superfluid hydrodynamics even when it should be valid, namely in the case of very strong pairing.

It seems therefore necessary to go one step further, which is the Hartree-Fock-Bogoliubov (HFB) theory \cite{Almirante24,Yoshimura24}, to reconcile the two approaches and solve this puzzle. If the periodicity of the system is taken into account (in contrast to the Wigner-Seitz approximation \cite{Pastore11} where only a single cell is considered), the HFB approach and its time-dependent extension (TDHFB) \cite{Yoshimura24} include the full information of the band structure. Moreover, unlike the BCS approximation (which includes only diagonal matrix elements of the gap), they should also be able to correctly reproduce the hydrodynamical behavior of the Cooper pairs in the limit of very strong pairing, as it was discussed in the context of cold atoms \cite{Grasso05,Tonini06}. The fact that the full HFB theory is needed and not only the simpler BCS approximation was demonstrated in the case of a toy model in Ref.~\cite{Minami22}.

In our previous work \cite{Almirante24}, we treated the slab phase in the HFB framework, finding a superfluid density slightly larger than the one expected from normal band theory \cite{Carter05}. Also, TDHFB calculations for the slab phase have been performed in \cite{Yoshimura24}, both with and without superfluidity. For comparable parameters, their results in the presence of superfluidity are similar to ours, while the ones without superfluidity are similar to those of the normal band theory. The observation that the superfluid density is higher than the density of ``free neutrons'' led the authors of \cite{Yoshimura24} to speak of ``anti-entrainment'' instead of entrainment. However, this statement depends on the definition of free neutrons, which is not unique. Fortunately, as pointed out in \cite{Chamel17,Pethick10}, the relevant microscopic parameter for the hydrodynamic description of the crust is the superfluid density $\rho_S$ and not the density of free neutrons. Hence, we will restrict ourselves to the evaluation of $\rho_S$, and we define the superfluid fraction $\rho_S/\bar{\rho}_n$ such that it refers only to the average neutron density $\bar{\rho}_n$ for which there is no ambiguity.

In the present work, we address this problem again by performing HFB calculations, at this time in the rod (``spaghetti'') phase, building up on our previous work on the slab (``lasagna'') phase \cite{Almirante24}. We will perform time independent calculations including a relative flow between the clusters and the superfluid component in a stationary way. This is possible since we treat neutrons as superfluid but protons as normal, implying that if there is a flow, the state of our system will depend only on the relative velocity between the cluster and the superfluid component (and not on two velocities as it would be the case if also protons were superfluid). Hence, a simple Galilean transformation is sufficient to retrieve a spatially periodic situation in spite of the flow.
\newline

In Sec.~\ref{sec:formalism}, we recall the interactions we use to construct the HFB matrix, the formalism for two-fluid hydrodynamics and the inclusion of a relative flow. In Sec.~\ref{sec:results}, results without and with flow are shown and discussed, together with a detailed comparison with results from normal band theory. Section \ref{sec:conclusion} contains the conclusions and perspectives. Details about the HFB formalism in a periodic lattice, the reciprocal lattice, and the numerical implementation are given in the Appendices.

\section{Formalism} \label{sec:formalism}
Since the formalism is essentially the same as in our previous work about the slab phase \cite{Almirante24}, we will limit ourselves in this section to a brief reminder and to the discussion of some extensions. For details please refer to our previous work. The formalism is kept general (3D)\, but the applications are performed for the rod (``spaghetti'') phase with $L$-periodicity in the $xy$-plane, for square and hexagonal lattices. Details that are specific to the 2D case are given in Appendices \ref{appA} and \ref{appB}.
\subsection{Hamiltonian}
In order to compute the microscopic properties of the inner crust of neutron stars, one has to deal with a system of superfluid neutrons and normal fluid protons arranged in a periodic lattice. Therefore, we perform HFB calculations for neutrons and Hartree-Fock (HF) ones for protons. (Our system contains also electrons, but they are assumed to form a constant uniform charge background, ensuring charge neutrality).

For the mean-field, we implement Skyrme-type energy-density functionals, from the SLy family and, as an extension of our previous work \cite{Almirante24}, also the BSk one. The difference between these two cases is in the coefficients related to the parameters $t_1$, $t_2$, which are taken density dependent in the BSk, including more parameters.

For simplicity, we neglect the spin-orbit term. Concerning the parametrizations, we use SLy4 \cite{Chabanat97} and BSk24 \cite{Chamel09,Chamel13A}. For the protons, we include the Coulomb energy, considering also the exchange term in the Slater approximatiom (the latter only in SLy4 since it is not included in the BSk24 parameter fit).

Taking the functional derivatives of the energy density with respect to number density $\rho_q$, kinetic energy density $\tau_q$ and momentum density $\jv_q$ (see \cite{Almirante24} for the relations between these and effective masses $m^\star_q$, mean-field potentials $U_q$ and the momentum dependent terms $\Jv_q$ required by Galilean invariance in Eq.~(\ref{eq:hmeanfield})), where $q = n,p$, one gets the mean-field Hamiltonian, which for each species reads in momentum space
\begin{equation}
\label{eq:hmeanfield}
     h_{\kv\kv'} =
     \kv\cdot\kv'\Big(\frac{\hbar^2}{2m^*}\Big)_{\kv-\kv'} \hspace{-1mm}+
     U_{\kv-\kv'}
     - (\kv+\kv') \cdot \Jv_{\kv-\kv'}\,.
\end{equation}
In order to consider a flow in our system, we replace in the HFB equations the mean-field Hamiltonian (\ref{eq:hmeanfield}) by
\begin{equation}
\label{eq:Hpv}
    h_{\kv\kv'}(\vv)=h_{\kv\kv'}
    - \hbar\kv\cdot\vv \delta_{\kv\kv'}\,.
\end{equation}
Due to the last term, there will be non-vanishing momentum densities $\jv_q$ (see \cite{Almirante24} for details on this construction). Notice that since protons are not superfluid, Galilean invariance ensures that $\vv$ will be the actual proton velocity.   
\vspace{3mm}\\
For the pairing field, we use a non-local interaction written in a separable form, namely
\begin{equation}
    V_{\kv_1\kv_2\kv_4\kv_3}^{\text{pair}} = -g
    \,
    f\Big(\frac{|\kv_1+\kv_2|}{2}\Big)
    f\Big(\frac{|\kv_3+\kv_4|}{2}\Big)
    \delta_{\kv_1-\kv_2,\kv_3-\kv_4}\,,
\end{equation}
The form factors $f(k)$ are taken to be Gaussians 
\begin{equation}
    f(k)=e^{-k^2/k_0^2}\,.
\end{equation}
The coupling constant $g$ and the Gaussian width $k_0$ had been fitted on the $V_{\text{low-}k}$ interaction in \cite{Martin14}. With the above interaction, the pairing gap will be non-local too, and in momentum space it will read
\begin{equation} \label{eq:gapk}
    \Delta_{\kv\kv'} = g \hspace{1mm}
    f\Big(\frac{|\kv+\kv'|}{2}\Big) \sum_{\pv\pv'}
    f\Big(\frac{|\pv+\pv'|}{2}\Big) \kappa_{\pv\pv'}
    \delta_{\kv-\kv',\pv-\pv'}\,,
\end{equation}
where $\kappa_{\pv\pv'}$ is the anomalous density matrix. One can also write the non-local pairing gap in Wigner (phase-space) representation as
\begin{equation} \label{eq:gapsep}
    \Delta(\Qv,\xv)= f(\Qv) \Delta_0(\xv)\,,
\end{equation}
with $\xv$ the Cooper pair c.o.m. position and $\Qv$ its relative momentum. Notice that since the form factor $f(k)$ is real, the above expression can be rewritten as
\begin{equation}
    \Delta(\Qv,\xv)= f(\Qv) |\Delta_0(\xv)| e^{i\phi(\xv)}\,,
\end{equation}
where $\phi$ is the phase of the pairing field. The separable pairing interaction makes the phase $\phi$ a function of the pair c.o.m. position only.

\begin{figure}
    \centering
    \includegraphics[width=8.5cm]{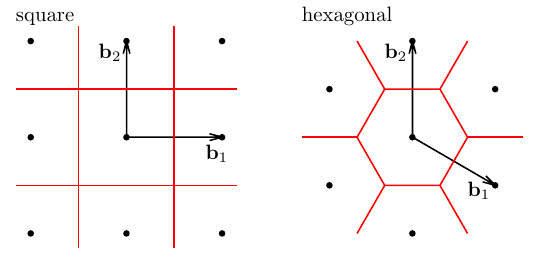}
    \caption{Reciprocal lattices for square (\textit{left}) and hexagonal (\textit{right}) lattice. Black points are the lattice points. The central square and hexagon delimited by the red lines are the respective first Brillouin zones. Reciprocal lattice primitive vectors $\bv_1$ and $\bv_2$ are the ones defined in (\ref{eq:ksquare}) and (\ref{eq:khexagon}).}
    \label{fig:replatt}
\end{figure}

We implement the above construction in momentum space, imposing Bloch boundary conditions. In practice, this means that the momentum $\kv$ (and analogously $\kv'$) can be split into the momentum $k_{\paral} = k_z$ parallel to the rods, a discrete lattice (the reciprocal lattice) in the $xy$ plane, and a continuous Bloch momentum $\kv_b$ that is restricted to the first Brillouin zone (BZ). In this representation, the matrices $h$ and $\Delta$ are diagonal in $k_z$ and $\kv_b$ and the diagonalization has to be performed on the reciprocal lattice. In Appendices \ref{appA} and \ref{appB} we discuss this in detail, but for the sake of clarity we show in Fig.~\ref{fig:replatt} the general structure of our momentum space for both explored lattices.

\subsection{Two-fluid flow}
In order to study the flow in systems in which superfluid and normal phases coexist, one can apply the formalism developed by Andreev and Bashkin \cite{Andreev75}. Since our protons are not superfluid, the Andreev-Bashkin relations for the particle currents $\rhov_q$ are reduced to
\begin{align} \label{AB4}
    \rhov_n &=(\rho_n-\rho_S)\vv_N +
    \rho_S \vv_S\,,\\
    \rhov_p &=\rho_p \vv_N\,,
\end{align}
where $\rho_n$ and $\rho_p$ are the number densities of neutrons and protons, $\vv_N$ is the velocity of the normal fluid, i.e., of the clusters, which coincides with the proton velocity $\vv$ we include in the mean-field Hamiltonian (\ref{eq:Hpv}), and $\rho_S$ and $\vv_S$ are the neutron superfluid density and neutron superfluid velocity, respectively.

In our inhomogeneous system the above relations make sense only at a ``coarse-grained'' scale. For this reason, $\rho_n$ and $\rho_p$ should be replaced with their cell averages $\bar{\rho}_n$ and $\bar{\rho}_p$. The average neutron superfluid velocity is given by the gradient of the coarse-grained phase $\bar{\phi}$ of the pairing field \cite{Pethick10}, i.e.,
\begin{equation} \label{eq:supvel1}
    \vvbar_S = \int_S \frac{dxdy}{S}
    \frac{\hbar}{2m}\nablav\phi\,,
\end{equation}
where $S$ is the surface of the lattice primitive cell and $\phi$ is the microscopic (not averaged) phase. It is not necessary to average over $z$ because the system is uniform in $z$ direction. Also, it should be noticed that $\rho_S$ is a tensor, i.e., it depends on the direction of the flow. Because of translational invariance in $z$ direction, it is clear that $\rho_{S,zz}=\bar{\rho}_n$, and $\rho_{S,xz} = \rho_{S,zx}=\rho_{S,yz} = \rho_{S,zy}=0$ for symmetry reasons. From now on, we will concentrate on the case that $\vvbar_S$ lies in the $xy$ plane, where the system has periodic inhomogeneities. In both square and hexagonal symmetry, $\rho_S$ can be shown to be isotropic in the $xy$ plane, i.e., $\rho_{S,xx}=\rho_{S,yy}$ and $\rho_{S,xy}=\rho_{S,yx} = 0$ \cite{Durel2018}. In the rest of this paper, $\rho_S$ will refer to the $\rho_{S,xx}$ ($ = \rho_{S,yy}$) component of the $\rho_S$ tensor.

Since we work in the frame in which the phase of the pairing field is periodic, Eq.~(\ref{eq:supvel1}) implies that the average superfluid velocity vanishes ($\vvbar_S=0$) and thus Eq.~(\ref{AB4}) becomes
\begin{equation} \label{supdens}
    \bar{\rhov}_n=(\bar{\rho}_n-\rho_S)\vv_N\,.
\end{equation}
With this relation, knowing densities and currents, we can directly infer the neutron superfluid density.

\section{Results} \label{sec:results}
\subsection{Density profiles and pairing gaps}
In our calculations, we will analyze nuclear matter under $\beta$-equilibrium. This condition implies that the chemical potentials satisfy
\begin{equation}
    \mu_n=\mu_p+\mu_e\,,
\end{equation}
where $\mu_n$ is fixed and $\mu_e$ is computed with the ultra-relativistic expression
\begin{equation}
    \mu_e = \hbar c (3\pi^2\bar{\rho}_e)^\frac{1}{3}\,.
\end{equation}
The electron density is determined requiring charge neutrality, and $\mu_p$ is readjusted after each HFB iteration to satisfy the $\beta$-equilibrium condition.

In principle, for given $\mu_n$, one should use the extension $L$ that minimizes the thermodynamic potential \cite{Martin15}, which is equivalent to minimizing the energy for given baryon density as done in \cite{Yoshimura24}. But the minimum is very flat and depends sensitively on details of the chosen interaction. In order to be able to compare with other calculations, e.g. \cite{Carter05}, we will consider different values for the cell extension $L$ in the range expected for the rod phase.
\begin{figure}
    \centering
    \includegraphics[width=8.5cm]{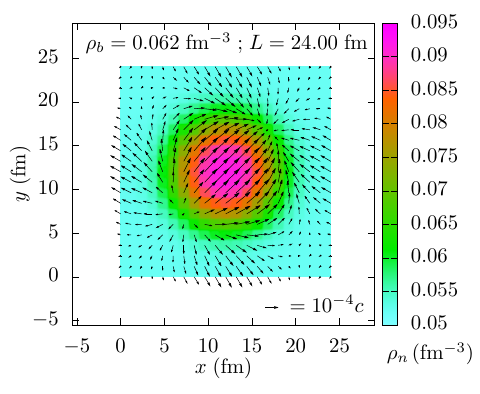}
    \includegraphics[width=8.5cm]{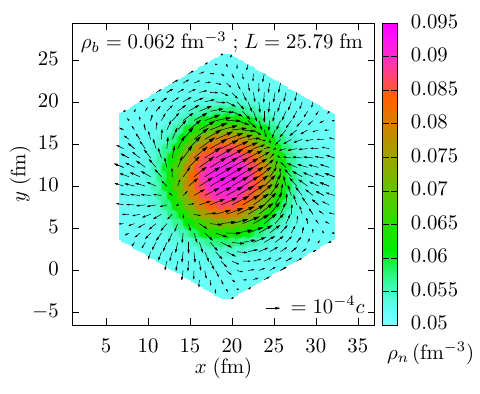}
    \caption{Neutron density (\textit{colors}) and velocity field (\textit{arrows}) in square (\textit{top}) and hexagonal (\textit{bottom}) lattice for SLy4, chemical potential $\mu_n=12$ MeV, baryon density $\rho_b=0.062$ fm$^{-3}$, cell surface $24\times24$ fm$^2$, velocity $\vv_N/c=(10^{-3},10^{-3})$ in the respective reciprocal basis. Notice that the cell extension refers to the distance between two clusters, thus the edge of the hexagon is $L/\sqrt{3}$.}
    \label{fig:densityvelocity}
\end{figure}

In Fig.~\ref{fig:densityvelocity}, we show neutron density profiles and velocity fields $\vv_n(\xv) = \rhov_n(\xv)/\rho_n(\xv)$ (the latter will be discussed in the next subsection) for the two kinds of lattices we analyze. The calculations were done with the SLy4 functional. The extension of the primitive cell is taken such that the surface for both cases is the same, namely
\begin{equation}
    S =  L_{\text{sq}}^2=\frac{\sqrt{3}}{2}L_{\text{hex}}^2\,.
\end{equation}
Density profiles and other relevant quantities such as effective mass and pairing field are practically identical. This is due to the fact that we are using the same interaction, chemical potential and cell volume and that the rods are well separated from each other and have a cylindrical shape.

Changing the interaction to BSk24, we can readjust the chemical potential in order to compensate to a large extent the different mean fields and have again the same baryon density. In this way we get comparable density profiles.
\begin{figure}
    \centering
    \includegraphics{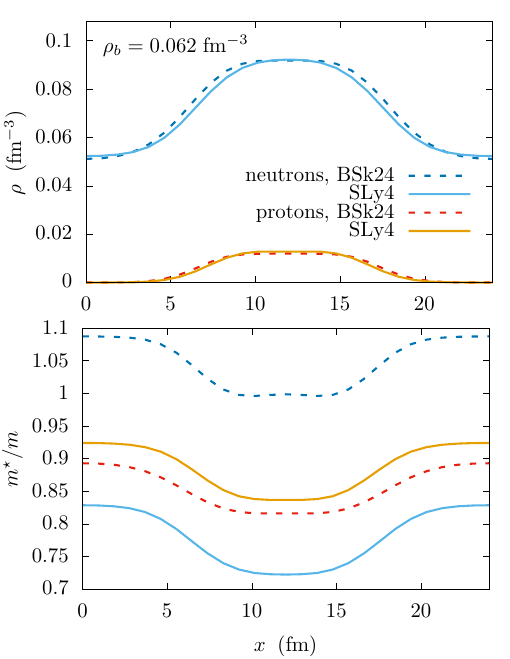}
    \caption{Square lattice section at $y=L/2$ of density profiles (\textit{top}) and effective masses (\textit{bottom}) for neutrons and protons. Results for interactions SLy4 and BSk24 are shown for the same $\rho_b=0.062$ fm$^{-3}$ and cell extension $L=24$ fm. The corresponding neutron chemical potentials are $\mu_n^{\text{SLy}}=12$ MeV and $\mu_n^{\text{BSk}}=10.62$ MeV.}
    \label{fig:densitymass}
\end{figure}
In the upper panel of Fig.~\ref{fig:densitymass}, one can see that, compared to SLy4, the neutron density in the BSk24 is slightly reduced in the gas and almost unchanged in the cluster, with slightly increased size of the latter. In the range of density were the rod phase is expected ($\rho_b\simeq0.06-0.07$ fm$^{-3}$), the main difference between the two functionals at the mean-field level is in the microscopic effective mass $m^\star$. In the lower panel of Fig.~\ref{fig:densitymass}, we show what we get for the ratio $m^\star/m$.
As it can be seen, in the neutron gas, our results are in agreement with the ones for pure neutron matter \cite{Duan24}, where at these densities $m^\star/m$ is expected to be less than unity for SLy4 and the opposite for BSk24.

\begin{figure}
    \centering
    \includegraphics{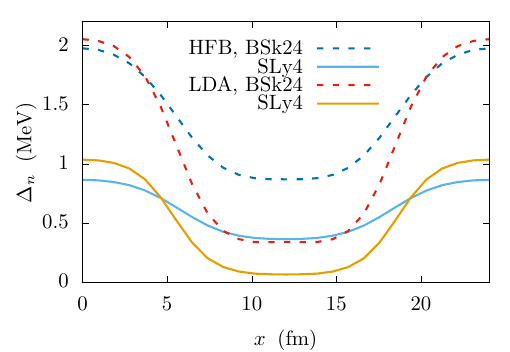}
    \caption{Square lattice section at $y=L/2$ of the neutron pairing gap at the local Fermi momentum as a function of the pair c.o.m. position for the same parameters as in Fig.~\ref{fig:densitymass}.}
    \label{fig:gap}
\end{figure}

Apart from the mean field we find a remarkable difference in the value of the pairing gap. This is mainly due to the difference in the effective mass, and we find it already at the Local-Density Approximation (LDA) level, which is the gap computed at the local density, effective mass and chemical potential, see \cite{Almirante24} for details. Our results for the HFB and LDA gaps are shown in Fig.~\ref{fig:gap}, both of them are taken at the local Fermi momentum, i.e. $Q=(3\pi^2\rho_n(\xv))^{1/3}$ in Eq.~(\ref{eq:gapsep}). Similar behaviors were found in \cite{Chamel10} for the 3D geometry when comparing the simpler BCS approximation with LDA gaps.
The better agreement with the LDA in the gas for the BSk24 pairing gap can be understood in terms of the coherence length $\xi$ of the Cooper pairs (as we analyzed it in our previous work \cite{Almirante24}): with increasing effective mass and pairing gap, $\xi$ becomes smaller, making the value of the gap closer to the LDA limit. Notice that in spite of this difference in the pairing gap, the densities in both cases are only slightly affected by the superfluidity, since the mean field potential remains at least one order of magnitude greater than the gap.

\begin{figure}
    \centering
    \includegraphics{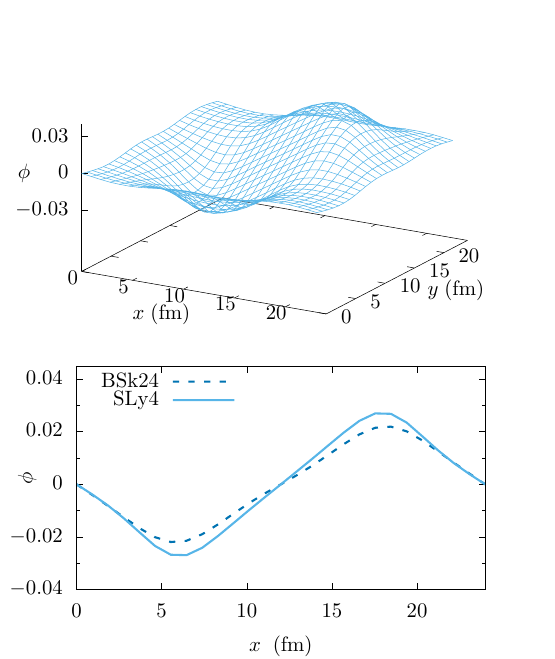}
    \caption{Phase of the pairing field for SLy4 in the square lattice (\textit{top}) and its section at $y=L/2$ (\textit{bottom}). For comparison, we show also BSk24 results in the bottom panel. The proton velocity is $\vv_N/c=(10^{-3},10^{-3})$ as in Fig.~\ref{fig:densityvelocity} and the other parameters are the same as in Fig.~\ref{fig:densitymass}.}
    \label{fig:phase}
\end{figure}

\subsection{Superfluid flow and phase of the gap}
Let us now turn to the main point of this paper, which is the case that there is a relative flow between clusters and superfluid neutrons. In Fig.~\ref{fig:densityvelocity}, we presented already our results for the neutron velocity fields in the reference frame in which the cluster moves and the superfluid carries in average no momentum.
Apart from some subtleties related to momentum-dependent terms in the Skyrme functional (for more details please refer to \cite{Almirante24}), this means that we are in the rest frame of the superfluid. As it can be seen also in the rod phase \cite{Almirante24}, the neutrons in the cluster move in the direction of $\vv_N$ (i.e., of the protons), while we find a counterflow outside the cluster. But since we are now in 2D, the velocity field has also a component in the direction perpendicular to $\vv_N$ which, depending on the position, points towards or away from the cluster. This complicated flow pattern can be understood more intuitively in the rest frame of the cluster, which one gets by boosting the whole image by $-\vv_N$. Then one finds that the neutrons flow slightly more slowly inside the cluster than in the gas, and that they prefer to go through the cluster rather than around it. However, we are not showing this figure because the deviation from a constant velocity field $-\vv_N$ is too small to be clearly seen (notice that the velocities in Fig.~\ref{fig:densityvelocity} are about one order of magnitude smaller than $|\vv_N|$).

In Fig. \ref{fig:phase}, we show the behavior of the phase of the pairing field. Our results are in qualitative agreement with the ones obtained in \cite{Martin16} in the framework of superfluid hydrodynamics, with the difference that in our case the cluster has no sharp surface. Remembering that the neutron velocity is (approximately) proportional to $\nablav\phi$, we see again the features that were already discussed above. In the lower panel of Fig.~\ref{fig:phase}, we compare in more detail the phase for the two interactions SLy4 and BSk24. We see that with BSk24 the slope of the phase (i.e., the neutron velocity) inside the cluster is slightly lower than with SLy4, i.e., with BSk24 less neutrons move together with the cluster, and as a consequence the superfluid density is slightly higher. While it is not surprising that a larger gap results in a larger superfluid density, the quantitative difference in superfluid densities is very small, as one can see from the numbers given in Table~\ref{betaequilibrium}. The reason for this is that our results are not very far from the maximum possible superfluid fraction given by the hydrodynamic picture, which should be valid in the limit of a very large gap, i.e., $\xi\ll L,R$ ($R$ being the cluster radius) and which predicts $\rho_S/\bar\rho_n \approx 0.97$ for the present parameters (cf. Fig.~10 of Ref.~\cite{Martin16}).

For completeness, let us mention that we tested different values for the velocity to check that we are in the linear regime (which ensures that our value for the superfluid fraction is velocity independent, see \cite{Almirante24}), and also different directions of the flow to check the isotropy of the response. Our results for a choice between these equivalent conditions are collected in Table~\ref{betaequilibrium}.  
\begin{table}
  \caption{\label{betaequilibrium} 
    Results for baryon density $\rho_b$, average neutron gap at the local Fermi momentum $\bar{\Delta}_n$, superfluid fraction $\rho_S/\bar{\rho}_n$, and proton fractions $Y_p$ for different chemical potentials $\mu_n$, cell extensions $L$, and interactions (SLy4 and BSk24).}
  \begin{ruledtabular}
    \begin{tabular}{ccccccc}
                         int.&$\mu_n$&$L$ &$\rho_b$&$\bar{\Delta}_n$&$\rho_S/\bar{\rho}_n$&$Y_p$\\
                             &(MeV)  &(fm)&(fm$^{-3}$)&(MeV)& &(\%)\\\hline
        \multirow{2}{*}{SLy4}&\multirow{2}{*}{12.00}&24&0.0619&0.755&0.945&3.24\\
                             &                   &27&0.0620&0.752&0.946&3.26\\\hline
        \multirow{2}{*}{SLy4}&\multirow{2}{*}{12.50}&24&0.0670&0.570&0.954&3.32\\
                             &                   &27&0.0666&0.588&0.956&3.26\\\hline
        \multirow{2}{*}{BSk24}&\multirow{2}{*}{10.62}&24&0.0620&1.734&0.955&3.15\\
                             &                   &27&0.0619&1.725&0.956&3.19\\\hline
        \multirow{2}{*}{BSk24}&\multirow{2}{*}{10.95}&24&0.0670&1.506&0.964&3.17\\
                             &                   &27&0.0667&1.512&0.965&3.18
    \end{tabular}
  \end{ruledtabular}
\end{table}
Notice that the presented results are obtained in the square lattice. We are not showing the hexagonal ones since at equal baryon density $\rho_b$ and cell surface $S$ they are practically identical (as discussed at the beginning of this Section).

\subsection{Comparison with normal band theory}
In order to compare our results with previous ones obtained in normal band theory by Carter and Chamel \cite{Carter05}, we implement the same physical conditions as in \cite{Carter05}. The authors take the mean field potential from the work of Oyamatsu \cite{Oyamatsu93,Oyamatsu94}, where neutron and proton densities are given together with an energy-density functional. Mean field potentials are obtained by taking the derivatives of the latter and performing a Gaussian smearing (to account for the finite range of the nuclear interaction). There is no effective mass and the spin-orbit term is neglected. Moreover, the authors perform the computation not self-consistently, i.e. the mean field is treated as a fixed external potential.

Hence, we perform HFB calculations for neutrons keeping the mean field potential from \cite{Oyamatsu94} fixed, while doing the self-consistency in the pairing channel. In this way we can directly compare our superfluid fraction $\rho_S/\bar{\rho}_n$ with the inverse of the mobility ratio $K_{\paral}/K_{\perp}$ from \cite{Carter05} (see \cite{Almirante24,Carter06} for the relation between superfluid density $\rho_S$ and mobility coefficient $K_{\perp}$). Results are collected in Table~\ref{chamelcomparison}. 
\begin{table}
  \caption{\label{chamelcomparison} Results for average neutron density $\rho_n$ and superfluid fraction $\rho_S/\bar{\rho}_n$ for different chemical potentials $\mu_n$, cell extensions $L$ and non-selfconsistent mean fields taken from Oyamatsu \cite{Oyamatsu94} (see text) and comparison with corresponding results from normal band theory \cite{Carter05}.}
  \begin{ruledtabular}
    \begin{tabular}{cccccc}
        $\mu_n$&$L$ &$\bar{\rho}_n$&$\bar{\rho}^\text{\cite{Carter05}}_n$&$\rho_S/\bar{\rho}_n$&$(\rho_S/\bar{\rho}_n)^{\text{\cite{Carter05}}}$\\
        (MeV)  &(fm)&(fm$^{-3}$)&(fm$^{-3}$)& & \\\hline
        {27.2461}&27.17&0.0590&0.0581&0.931&0.681\\
        {28.7422}&25.77&0.0641&0.0630&0.936&0.756\\
        {30.1797}&24.62&0.0692&0.0678&0.943&0.825\\
        {31.2812}&23.97&0.0732&0.0716&0.949&0.872
    \end{tabular}
  \end{ruledtabular}
\end{table}
There is a small difference in the neutron density due to the presence of the pairing gap in our calculation. We find that neutron density in the gas is the same and the cluster size is comparable, while in the cluster our neutron density is slightly increased (about $5\%$). This is reflected in the average neutron density.
\newline
As it can be seen, our superfluid fraction is much larger than in \cite{Carter05}. Notice that our results are velocity independent because we are in the linear regime.

\begin{figure}
    \centering
    \includegraphics{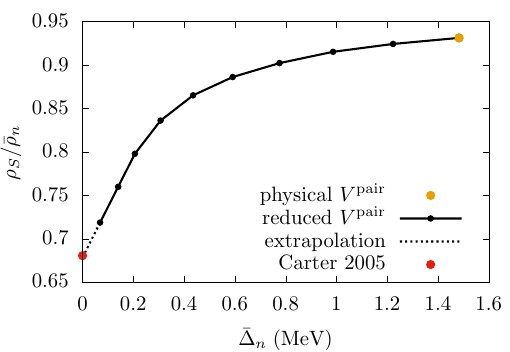}
    \caption{Neutron superfluid fraction $\rho_S/\bar{\rho}_n$ as a function of the average gap $\bar{\Delta}_n$. Black points are obtained by varying the pairing coupling constant in the interval $g/g_{\text{phys}}\in[0.55,0.95]$, other parameters are as in the first line of Table~\ref{chamelcomparison}. The yellow point is our result at the physical $V^{\text{pair}}$. The red point is the result from \cite{Carter05}. The black dashed line is a linear extrapolation from our results to $\bar{\Delta}_n=0$.}
    \label{fig:reduced_pairing}
\end{figure}

In order to understand where is the difference between the two computations, one can make the following numerical study. The results in \cite{Carter05} are obtained through a relation that can be viewed as the weak-coupling limit of an expression obtained in the BCS approximation \cite{Carter05A,Chamel17}. Thus, in order to bridge between our results and those of the normal band theory, we try to reproduce the weak-coupling limit. This can be done by reducing artificially the pairing coupling constant $g$. Our results are collected in Fig.~\ref{fig:reduced_pairing}. Notice that decreasing the pairing gap implies also a decrease in the critical velocity where pair breaking sets in, thus the results in Fig.~\ref{fig:reduced_pairing} are obtained with a cluster velocity $\vv_N/c=(10^{-4},10^{-4})$, which is sufficiently small to ensure the linear regime for all explored values of the gap. As it can be seen, the result from \cite{Carter05} appears to be the limit for $\bar{\Delta}_n=0$ of our more general treatment. Our results confirm earlier suspicions \cite{Martin16,Minami22} that doing the calculation in normal band theory as in \cite{Carter05} leads to an underestimation of the superfluid fraction.

Moreover, we see in Fig.~\ref{fig:reduced_pairing} that, as a function of the gap, the superfluid fraction rises first very quickly and then seems to approach a limiting value, which corresponds presumably to the superfluid hydrodynamics result. The fact that the curve is already quite flat at the physical pairing strength explains why, as we have seen in the preceding subsection, the precise value of the pairing gap affects only weakly the superfluid fraction. In \cite{Carter05A} it was said (for the 3D case) that inclusion of pairing in the BCS approximation
does not lead to a significant change of the superfluid fraction compared to the normal band theory. 
Hence, it seems to be important that pairing is included within HFB and not only within the BCS approximation. Although both theories agree in the weak-coupling limit, only the HFB theory is able to correctly reproduce also the hydrodynamic limit at strong coupling.

\section{Conclusion and perspective} \label{sec:conclusion}
In this work we investigated the superfluid features of the periodic rod phase in $\beta$-equilibrium at baryon density $\rho_b\simeq0.06-0.07$\hspace{1mm}fm$^{-3}$ and cell extension $L\simeq24-27$\hspace{1mm}fm, as it is expected in the inner crust of neutron stars. This has been done performing HFB calculations with Bloch boundary conditions, making use of an energy density functional for the mean-field and of a non-local separable potential fitted on $V_{\text{low-}k}$ for the pairing field.

We implemented two different lattices, namely square and hexagonal ones. At equal interaction, baryon density, and cell surface, we found that they give practically the same results for neutron and proton density profiles, neutron gap, and superfluid density.

For the mean fields, we implemented two different energy density functionals, namely SLy4 and BSk24. Compared to SLy4, we found that at equal baryon density and cell surface the neutron density in the gas is slightly decreased and the cluster size is slightly increased in BSk24. More importantly, the neutron effective masses are larger in BSk24 than in SLy4, in agreement with results for pure neutron matter \cite{Duan24}. Therefore, the pairing gap is much larger in the BSk case.

The most intriguing results concern the superfluid dynamics. We include a relative stationary flow between the clusters and the surrounding superfluid neutron gas, in a linear regime for the relative velocity. We found that in the superfluid rest frame (in the sense $\vvbar_S = 0$), outside the cluster there is in average a counterflow of neutrons in the direction opposite to the one in which the cluster moves, while inside the cluster they flow in the same direction as it. This is also clear from the periodicity of the phase of the pairing field. Furthermore, there is also some motion in the perpendicular direction which comes from the fact that, in the rest frame of the cluster, the neutrons prefer to flow through it. In fact, our velocity fields are comparable with the ones from superfluid hydrodynamics \cite{Martin16}, and actually we found for both explored functionals a superfluid fraction $\rho_S/\bar{\rho}_n\simeq95\%$ which is almost as large as the prediction of superfluid hydrodynamics, which is $\simeq 97\%$.

This is corroborated also by our comparison with results obtained from normal band theory by Carter and Chamel \cite{Carter05}. For the purpose of a quantitative comparison, we implemented the same mean-field potentials as \cite{Carter05} (taken from \cite{Oyamatsu93,Oyamatsu94}) but including the pairing channel self-consistently.
Our superfluid fraction obtained in this way is much larger than in \cite{Carter05} and comparable to the one we got in the fully self-consistent treatment (whose neutron densities are close to those of \cite{Oyamatsu93}). However, we can reproduce the results of normal band theory to very good precision if we go into the weak-coupling limit by artificially reducing the pairing strength. This shows that HFB theory can interpolate between the normal band theory which is only valid in the weak-coupling limit, and superfluid hydrodynamics which is only valid in the limit of very strong pairing.

In summary we found that, at least in the rod phase, the superfluid density is larger than predicted previously by normal band theory calculations and closer to the prediction of superfluid hyrodynamics. Our aim for the future is to perform the calculations as done in this work for the 3D crystal phase of the inner crust, which is the most extended one and thus it would have the strongest astrophysical impact.
\appendix
\section{HFB with Bloch boundary conditions in 2D} \label{appA}
We want to solve the HFB equations in a 2D periodic lattice. In order to do this, one can introduce the $L$-periodicity condition in the $xy$-plane (details about the particular choices taken in this work are given in Appendix~\ref{appB}) for the basic quantities, namely density and anomalous density, i.e.,
\begin{gather}
\langle \psi^\dagger_\uparrow(\rv'+\av_i) \psi_\uparrow(\rv+\av_i) \rangle 
  = \langle \psi^\dagger_\uparrow(\rv')\psi_\uparrow(\rv) \rangle\,,
  \\
\langle \psi_\downarrow(\rv'+\av_i) \psi_\uparrow(\rv+\av_i) \rangle 
  = \langle \psi_\downarrow(\rv') \psi_\uparrow(\rv) \rangle\,,
  \label{perd}
\end{gather}
where $\av_i$ are the primitive vectors of the direct lattice. Then one also requires translational invariance in $z$ direction, i.e., $\forall \lambdavpar = \lambda\zvhat$
\begin{gather}
\langle \psi^\dagger_\uparrow(\rv'+\lambdavpar) \psi_\uparrow(\rv+\lambdavpar) \rangle 
  = \langle \psi^\dagger_\uparrow(\rv') \psi_\uparrow(\rv) \rangle\,,
  \\
\langle \psi_\downarrow(\rv'+\lambdavpar) \psi_\uparrow(\rv+\lambdavpar) \rangle 
  = \langle \psi_\downarrow(\rv') \psi_\uparrow(\rv) \rangle\,.
  \label{homo}
\end{gather}
For the anomalous density matrix one can write
\begin{equation}
\langle c_{-\pv'\downarrow} c_{\pv\uparrow} \rangle
 = \int \! d^3 r\, d^3r' e^{-i\pv\cdot\rv}\, e^{i\pv'\cdot\rv'} \langle \psi_\downarrow(\rv') \psi_\uparrow(\rv) \rangle
   \,.
   \label{eq:ccavg}
\end{equation}
Performing a change in the integration variables $(\rv,\rv') \to (\rv+ \av_i + \lambdavpar,\rv' + \av_i + \lambdavpar)$ and using Eqs.~(\ref{perd}) and (\ref{homo}), one obtains
\begin{multline}
\langle c_{-\pv'\downarrow} c_{\pv\uparrow} \rangle 
  =\int\! d^3 r\, d^3 r'\, e^{-i\pv\cdot\rv}\, e^{i\pv'\cdot\rv'} \langle \psi_\downarrow(\rv') \psi_\uparrow(\rv) \rangle
  \\
  \times e^{-i(p_i-p'_i)L} e^{-i(p_{\paral}-p'_{\paral})\lambda}\,,
  \label{eq:ccavgshifted}
\end{multline}
where $p_i$ are components of the momentum $\pv$ in the basis $\{\hat{\bv}_1,\hat{\bv}_2\}$ of the reciprocal lattice (see Appendix \ref{appB}) and $p_{\paral} = p_z$ (and analogously for $p'_i$ and $p'_{\paral}$). Combining Eqs.~(\ref{eq:ccavg}) and (\ref{eq:ccavgshifted}), one finds that the momentum labels must satisfy the following conditions: $p_1-p'_1=\frac{2\pi}{L}\nu_1$, $p_2-p'_2=\frac{2\pi}{L}\nu_2$  with $\nu_1,\nu_2\in\mathbb{Z}$ and $p_{\paral}-p'_{\paral} = 0$. Hence, the momentum dependence of the matrix element can be written as
\begin{equation}
    \langle c_{-\pv'\downarrow} c_{\pv\uparrow} \rangle =
    \delta_{\pv_b,\pv'_b} \delta_{p_{\paral},p'_{\paral}}
    \langle c_{-\nv'\downarrow}(-\pv_b,-p_{\paral}) c_{\nv\uparrow}(\pv_b,p_{\paral}) \rangle\,,
\end{equation}
where we rewrote the momentum components in 1 and 2 directions as a sum of an integer multiple of $\frac{2\pi}{L}$ and the Bloch momentum defined in the first BZ, namely $p_1=\frac{2\pi}{L}n_1+p_{b1}$, $p_2=\frac{2\pi}{L}n_2+p_{b2}$, where $n_1,n_2\in \mathbb{Z}$ with $\nv=(n_1,n_2)$ and $p_{b1}, p_{b2}\in (-\frac{\pi}{L},\frac{\pi}{L}]$ with $\pv_b=(p_{b1},p_{b2})$.

For the normal density matrix one can proceed in a completely analogous way and gets
\begin{equation}
    \langle c^\dagger_{\pv'\uparrow} c_{\pv\uparrow} \rangle =
    \delta_{\pv_b,\pv'_b} \delta_{p_{\paral},p'_{\paral}}
    \langle c^\dagger_{\nv'\uparrow}(\pv_b,p_{\paral}) c_{\nv\uparrow}(\pv_b,p_{\paral}) \rangle\,.
\end{equation}

As a consequence of these relations, our HFB matrix is diagonal in $p_{b1}$, $p_{b2}$, and $p_{\paral}$. Moreover, in our HFB matrix there is no dependence on the sign of the parallel momentum, i.e., it depends only on $p_{\paral}^2=p_z^2$

Summarizing, for each triple $(p_{b1},p_{b2}, p_{\paral})$ we have an HFB matrix with indices $n_1,n_2,n'_1,n'_2$, namely
\begin{equation}\label{eq:HFBH}
 \calH =
 \begin{pmatrix}
     h-\mu & -\Delta \\
     -\Delta^\dagger & -\bar{h}+\mu
 \end{pmatrix} ,
\end{equation}
where $h$ is the mean-field Hamiltonian (including the term $-\pv\cdot \vv$), $\Delta$ is the pairing field and $\bar{h}_{\kv \kv'}=h_{-\kv' -\kv}$.
We diagonalize it, obtaining quasi-particle energies $E_\alpha(\pv_b,p_{\paral})$ and eigenvectors $(U^*_{\alpha \nv}(\pv_b,p_{\paral}),-V^*_{\alpha \nv}(\pv_b,p_{\paral}))$. In terms of these, the normal and anomalous density matrices are expressed as
\begin{gather}
    \rho_{\nv\nv'}(\pv_b,p_{\paral})=\sum_{E_\alpha>0} V_{\nv' \alpha}^*(\pv_b,p_{\paral}) V_{\nv \alpha}(\pv_b,p_{\paral})\,,\\
    \kappa_{\nv\nv'}(\pv_b,p_{\paral})=\sum_{E_\alpha>0} U_{\nv \alpha}^*(\pv_b,p_{\paral}) V_{\nv' \alpha}(\pv_b,p_{\paral})\,,
\end{gather}
with $E_\alpha = E_\alpha(\pv_b,p_{\paral})$. Then, using the short-hand notation
\begin{equation}
    \sum_{\pv_b p_{\paral}} = \frac{1}{4\pi^3} \int_0^\infty dp_{\paral}\,\int_{\text{BZ}} dp_{b1}\hspace{1mm}dp_{b2}\,,
\end{equation} 
one can compute the densities and pairing field as
\begin{equation} \label{eq:dens1D}
    \rho(\xv) = 2\sumpbppar
    \sum_{\nv \nv'} e^{i\frac{2\pi}{L}(\nv-\nv')\cdot\xv}\,\rho_{\nv \nv'}(\pv_b,p_{\paral})\,,
\end{equation}
\begin{multline}
    \tau(\xv) = 2\sumpbppar
    \sum_{\nv \nv'}e^{i\frac{2\pi}{L}(\nv-\nv')\cdot\xv}
    \rho_{\nv \nv'}(\pv_b,p_{\paral})\\
     \times\Big( \Big(\frac{2\pi}{L}\nv+\pv_b\Big)\cdot\Big(\frac{2\pi}{L}\nv'+\pv'_b\Big)+p_
     {\paral}^2\Big)\,,
\end{multline}
\begin{multline}
    \jv(\xv) = 2
    \sumpbppar
    \sum_{\nv \nv'}  e^{i\frac{2\pi}{L}(\nv-\nv')\cdot\xv}
    \rho_{\nv \nv'}(\pv_b,p_{\paral})\\
    \times\Big(\frac{\pi}{L}(\nv+\nv')+\pv_b\Big) \,,
\end{multline}
\begin{multline} \label{eq:gapkn}
    \Delta_{\nv \nv'}(\kv_b,k_{\paral})= g f_{\nv+\nv'}(\kv_b,k_{\paral}) \sum_{\vect{m} \vect{m}'} \delta_{\nv-\nv',\vect{m}-\vect{m}'}\\
    \times\sumpbppar
    f_{\vect{m}+\vect{m}'}(\pv_b,p_{\paral}) \kappa_{\vect{m}\vect{m}'}(\pv_b,p_{\paral})\,,
\end{multline}
where
\begin{equation}
    f_{\nv+\nv'}(\kv_b,k_{\paral}) =
    \exp\bigg(-\frac{(\frac{\pi}{L}(\nv+\nv')+\kv_b)^2+k_{\paral}^2}{k_0^2}\bigg)\,.
\end{equation}

All of this is also true for the HF case, with the simplification that there are no anomalous density and pairing field. Thus, instead of the HFB matrix, only the mean-field Hamiltonian $h_{\nv \nv'}(\pv_b,p_{\paral})$ needs to be diagonalized. Denoting its eigenvalues and eigenvectors $\epsilon_\alpha(\pv_b,p_{\paral})$ and $V_{\alpha \nv}(\pv_b,p_{\paral})$, the density matrix reads then
\begin{equation} \label{eq:densmatHF}
    \rho_{\nv \nv'}(\pv_b,p_{\paral})=\sum_{\epsilon_\alpha<\mu} V_{\nv' \alpha}^*(\pv_b,p_{\paral}) V_{\nv \alpha}(\pv_b,p_{\paral})\,.
\end{equation}
%
\section{Reciprocal lattice and Brillouin zone} \label{appB}
In Appendix~\ref{appA} we discussed the general formalism for HFB with Bloch boundary conditions in 2D, however it is important to notice that while in the 1D case the above construction uniquely defines the HFB equations, this is no longer true in 2D. This is due to the fact that in 2D different kinds of lattices can be defined and thus one has to inform the problem of the specific lattice geometry. In general, the basis vectors of the direct lattice can be arbitrary vectors in the $xy$ plane, having a certain angle between them (thus not necessarily a trivial scalar product), and consequently the reciprocal vectors may be not aligned with them. Since we are working in momentum space, it is important to construct the reciprocal lattice and the BZ in the right way.

For what concerns the reciprocal lattice, one can find its primitive vectors using the well-known relation \cite{Ashcroft}
\begin{equation} \label{eq:reclat}
    \bv_i\cdot\av_j=2\pi\delta_{ij}\,,
\end{equation}
where $\av_i$ is a primitive vector of the direct lattice, and we define the corresponding basis vectors as $\hat{\bv}_i = \bv_i/|\bv_i|$. Notice that this is the only condition necessary for the construction in Appendix~\ref{appA}, since it allows one to go from Eq.~(\ref{eq:ccavg}) to Eq.~(\ref{eq:ccavgshifted}) for any kind of lattice using the corresponding reciprocal vectors as a basis for momentum space.

For the square lattice one has
\begin{align} \label{eq:ksquare}
    \begin{cases}
    \av_1=L \xvhat,&
    \bv_1=\frac{2\pi}{L}\xvhat,\\
    \av_2=L\yvhat,&
    \bv_2=\frac{2\pi}{L}\yvhat,
    \end{cases}
\end{align}
while for the hexagonal one
\begin{align} \label{eq:khexagon}
    \begin{cases}
    \av_1=L\xvhat,&
    \bv_1=\frac{2\pi}{L}
    \big(\xvhat-\frac{1}{\sqrt{3}}\yvhat\big),\\
    \av_2=L\big(\frac{1}{2}\xvhat+\frac{\sqrt{3}}{2}\yvhat\big),&
    \bv_2=\frac{2\pi}{L}\big(\frac{2}{\sqrt{3}}\yvhat\big),
    \end{cases}
\end{align}
where $L$ is the periodic extension which is equal in both directions for these particular lattices.

Since we want to do numerics, our momentum space will be finite. Thus, in order to be consistent, also the bounds of our reciprocal lattice must have the symmetry of the reciprocal lattice itself. For the square lattice, this is trivial since it is enough to number a grid on the square. For the hexagonal one, we instead start from the numbered rhombus grid, then we shift the points lying in the far corners in order to fulfill 60$^\circ$ symmetry.

Once one has constructed the reciprocal lattice, one has to define the BZ such that it has again the same symmetries \cite{Ashcroft}. In our cases, this means that for 90$^\circ$ symmetry one has a square BZ, while for 60$^\circ$ symmetry one has an hexagonal BZ. A schematic view is given in Fig.~\ref{fig:replatt}. 
\section{Numerical details}
As shown in Appendix \ref{appA}, our problem is diagonal in the Bloch momenta $\kv_b$ and in the parallel momentum $k_{\paral}$, thus we deal with an HFB matrix in the integer momenta $\nv,\nv'$. Since we want to perform the self-consistent calculations on a machine, the problem has to be discretized. We choose to divide the cell extension $L$ in the two directions such that $L=N\Delta x_i$ with $i=1,2$. This naturally introduces a cutoff in the integer momenta
\begin{equation}
    \Lambda=\frac{2\pi}{L}N \quad {\Rightarrow}\quad
    n_1,n_2\in (-N/2,N/2]
\end{equation}
(before the reshuffling of the rhombus lattice into the hexagonal shape mentioned at the end of Appendix~\ref{appB}).
As a consequence, the HFB matrix has dimension $2N^2\times2N^2$ (the HF matrix instead only $N^2\times N^2$). In this way, we can access the first $N^2$ bands for both neutrons and protons. In this work, we choose $N=26$.

For each combination $(\kv_b,k_{\paral})$, a diagonalization of both the HFB and the HF matrices in momentum space is performed. The relevant quantities to construct the matrices are computed in coordinate space and then transformed through Fast Fourier Transform (FFT).

For $\kv_b$ of neutrons, we take $N_b^{\text{sq}}=64$ points in the first Brillouin zone of the square lattice, while $N_b^{\text{hex}}=48$ for the hexagonal lattice. This difference in the number of points is due to the different kind of integration performed. For the square lattice we use Gauss-Legendre points, while for the hexagonal one we use the rules from Lyness \cite{Lyness77}.

For $k_{\paral}$ of neutrons, we introduce a cutoff $\Lambda_{\paral}^{(n)}=2$ fm$^{-1}$, which is sufficiently large compared to the parameter $k_0$ in the pairing interaction, such that the pairing gap does not depend on it. In the interval $[0,\Lambda_{\paral}^{(n)})$, we use $N_{\paral}=120$ points. This number multiplied by the number of points in the Brillouin zone should be a multiple of the available number of cores of the machine, since we perform in parallel the diagonalizations for each combination $(\kv_b,k_{\paral})$.

With this choice, the average grid spacing in the parallel direction is approximately $\Delta k_{\paral}=\Lambda_{\paral}^{(n)}/N_{\paral}\approx 0.017\,\text{fm}^{-1}$. However, notice that also for $k_{\paral}$ integration we use Gauss-Legendre points.

In the cases in which the average neutron gap at the local Fermi momentum is very small, $\bar{\Delta}_n < 0.3$ MeV, encountered when computing the results for Fig.~\ref{fig:reduced_pairing}, we increase the number of integration points to $N_b^{\text{sq}}=144$ and $N_{\paral}=240$.

For protons, since they are strongly confined, only a few bands are occupied and they are flat. As an approximation one could replace the integration over the Brillouin zone with the product between the BZ surface and the integrand computed in $\kv_b=0$.

Unfortunately performing the calculation in this way gives rise to spurious results in the currents. This is a discretization effect coming from the asymmetry in the integer momenta bounds. Taking as example the square lattice, one has in the two directions $n_i\in(-N/2,N/2]$. If the $N/2$ mode has some contribution to the current (although it should be small), computing it in $\kv_b=0$ would give a term that, even when there should be no current, will not be cancelled by the corresponding term in $-N/2$. In order to avoid this kind of effects, we choose for both protons and neutrons integration points in the Brillouin zone such that none of their components is zero and, for each integration point, there are also those obtained from it by flipping one by one the signs of its components. Then, if one of the integer momentum components $n_i=N/2$, the BZ is split into two parts: for $k_{bi}<0$, we keep $n_i=N/2$, while for $k_{bi}>0$, we replace it by $n_i=-N/2$. In this way parity is respected at the BZ level and there is no spurious effect.

For protons we take $N_b=4$ points in the Brillouin zone, namely $\kv_{b1} = (\frac{\pi}{2L},\frac{\pi}{2L})$, $\kv_{b2} = (\frac{\pi}{2L},-\frac{\pi}{2L})$ and the couple following from them by $\kv_{b3,4} = -\kv_{b1,2}$.

The $k_{\paral}$ integration for protons is performed with equidistant points with cutoff $\Lambda_{\paral}^{(p)}=2$ fm$^{-1}$ (it is sufficient that it is larger than the maximum Fermi momentum). With $N_{\paral}=6\times240$ (again chosen according to the available number of cores), the corresponding spacing is $\Delta k_{\paral}\approx 0.0014$ fm$^{-1}$. This fine spacing is needed because the proton distribution contains a step function.

For all the choices we discussed about integration points and cutoffs, we have verified that if we increase them the self-consistent calculations converge to the same results. Our convergence criterion is defined such that for each point in the cell
\begin{align}
    |\rho^{(m)} - \rho^{(m+1)}| &< |\rho^{(m+1)}|\times10^{-4},\nonumber\\
    |j^{(m)} - j^{(m+1)}| &< |j^{(m+1)}|\times10^{-4},\nonumber\\
    |\Delta_0^{(m)} - \Delta_0^{(m+1)}| &< |\Delta_0^{(m+1)}|\times10^{-4},
\end{align}
where $\rho^{(m)}$ is the result of the $m$-th iteration etc.

In order to speed up the convergence, we use Broyden's modified method as discussed in \cite{Baran08} and already used in the HFB description of the slab phase in \cite{Almirante24,Yoshimura24}. Our Broyden vector has dimension $9 N^2 + 2$ and it is defined as ($U_n(\xv)$, $U_p(\xv)$, $\hbar^2/2m^*_n(\xv)$, $\hbar^2/2m^*_p(\xv)$, $\Jv_n(\xv)$, $\Jv_p(x)$, $F_{\nv-\nv'}$, $\mu_n$, $\mu_p$). $F_{\nv-\nv'}$ is the gap defined in Eq.~(\ref{eq:gapkn}) without the factor $gf_{\nv+\nv'}(\kv_b,k_{\paral}$). We performed the method using the results of $M=3$ previous iterations and a mixing coefficient $\alpha=0.7$.

\bibliography{refs}
\end{document}